# Autonomous Experiments in Scanning Probe Microscopy and Spectroscopy: Choosing Where to Explore Polarization Dynamics in Ferroelectrics

Rama K. Vasudevan,[1,a] Kyle Kelley,[1] Jacob Hinkle[2], H. Funakubo,[3] Stephen Jesse,[1] Sergei V. Kalinin,[1,b] Maxim Ziatdinov[2]

[1] The Center for Nanophase Materials Sciences, Oak Ridge National Laboratory, Oak Ridge, TN 37831
[2] Computational Sciences and Engineering Division, Oak Ridge National Laboratory, Oak Ridge, TN 37831
[3] Department of Material Science and Engineering, Tokyo Institute of Technology, Yokohama, 226-8502, Japan

**Abstract**

Polarization dynamics in ferroelectric materials are explored *via* the automated experiment in Piezoresponse Force Microscopy/Spectroscopy (PFM/S). A Bayesian Optimization (BO) framework for imaging is developed and its performance for a variety of acquisition and pathfinding functions is explored using previously acquired data. The optimized algorithm is then deployed on an operational scanning probe microscope (SPM) for finding areas of large electromechanical response in a thin film of $PbTiO_3$, with results showing that with just 20% of the area sampled, most high-response clusters were captured. This approach can allow performing more complex spectroscopies in SPM that were previously not possible due to time constraints and sample stability. Improvements to the framework to enable the incorporation of more prior information and improve efficiency further are modeled and discussed.

[a] vasudevanrk@ornl.gov
[b] sergei2@ornl.gov

Physical imaging methods based on scanning probes or electron and ion beams have become the mainstay of modern science, finding applications in the fields ranging from condensed matter and quantum physics, materials science, chemistry, biology, and medicine.[1-3] Despite the extreme variability of the imaging methods in terms of the probes, interaction mechanisms, and the type of information they provide, the nature of the imaging process based on the rectangular scanning grids remains the same. At the same time, it is well understood that the information of interest in most cases tends to be concentrated in a small number of spatial locations that are often associated with specific microstructural elements or imposed by non-local boundary effects. These considerations, combined with recent notable advances in the field of autonomous vehicles and robotics, have stimulated intensive discussion of the opportunities opened by the automated experiment in microscopy, with concepts ranging from tuning of specific parameters or regions of image space[4] to fully automated self-driving microscopes[5-7] being proposed.

Recent advancements in autonomous experiments have threaded through a spectrum of materials characterization[8-12] and synthesis techniques.[13-20] However, despite much enthusiasm over the last several years, the experimental progress has been limited in the field of microscopy. The reasons for this dearth of experimental implementation are manyfold. From an engineering viewpoint, most commercial microscopes utilize closed controllers, and enabling externally controlled operations necessitates access to the instrumentation hardware at a low-level. Even more significantly, the linearity of the scanning process often relies on the proprietary algorithms optimized for specific platforms. This is particularly the case for automated experiment (AE) methods that alter the probe motion trajectory, as opposed to techniques that utilize feedback-controlled signal during classical scanning.[21] Secondly, automated experiment requires specific algorithms that define the scanning trajectory and probe parameters (*e.g.* bias in SPM) based on previously detected signals, often requiring substantial computational work (*e.g.*, *via* "edge computing" devices). Finally, while considerably less obvious, automated experiment does not universally offer advantages compared to the classical experimental paradigm and requires an understanding about the cases in which the growing complexity of the imaging process is compensated by increases in data acquisition rates.

Previously, we have demonstrated control of the probe trajectory in scanning probe[22, 23] and electron[24, 25] microscopy (EM), as well as simple image-based feedback in EM.[26] Similarly, we have developed a Bayesian Optimization (BO) based framework for the reconstruction of the hyperspectral images[27] and applied it to the exploration of the low-dimensional parameter spaces of the theoretical models.[28, 29] Here, we develop the integrated workflow based on the combination of these two concepts and introduce the instrument-specific BO workflows that leverage the information explored (acquisition function) and scanning sequence (pathfinder function). This approach is implemented on a SPM with ready access to a bank of graphical processing units (GPUs) to explore polarization switching in multidomain ferroelectric materials.

The heart of AE in imaging is the sequential exploration of the image space, in our case the 2D image plane, with the probe. Upon evaluating the performance of the specific measurement sequence and given the instant and past observation, the algorithm seeks to establish the next location which offers the maximum gain of some pre-defined metric. Repeating this process ideally allows us to collect the maximal amount of information in a much smaller number of steps compared to classical grid-based scanning. While simple in principle, this workflow necessitates establishing several well-defined concepts, namely the nature of the signal of interest, and the balance between information (exploration) and discovery of regions with specific behaviors (exploitation). Furthermore, in the experimental setup, the motion of the probe is associated with instrument-specific times, necessitating balancing the expected information gain with the acquisition times.

At the center of the AE approach developed here is the Gaussian Process (GP) based BO framework. GP refers to a universal function approximator, which can be used for reconstructing a function $f(\mathbf{x})$ over parameter space $\mathbf{x}$ given the observations $y_i$ at specific values $\mathbf{x}_i$. It is a non-parametric method, where the reconstruction is performed under the most general assumptions about function $f(\mathbf{x})$. Specifically, it is assumed that the observations of $f_i$ represent noisy measurements of the function, $y_i = f(\mathbf{x}_i) + \epsilon$, where $\epsilon$ is Gaussian noise, and that the values of the function are related through a kernel function. The functional form of the kernel is chosen prior to the experiment, and kernel (hyper)parameters are determined self-consistently from the observations ($f_i$, $\mathbf{x}_i$).

The unique aspect of GP methods compared to other interpolation approaches is that GP analysis also provides quantified uncertainty, *i.e.* function $\sigma(\mathbf{x})$ determined over the same parameter space $\mathbf{x}$ that defines the standard de*via*tion of the expected values $f(\mathbf{x})$. This naturally allows extending the GP approach towards automated experiments. Here, after the $n$ initial measurements ($y_1, y_2, ..., y_n$) at locations ($\mathbf{x}_1, \mathbf{x}_2, ..., \mathbf{x}_n$) the function $f(\mathbf{x})$ and its uncertainty $\sigma(\mathbf{x})$ are reconstructed, and the location with maximal uncertainty, argmax($\sigma(\mathbf{x})$) is chosen for a subsequent measurement. In this regime, AE minimizes the uncertainty over the predictions, corresponding to a purely exploratory strategy. In the BO methods, the exploration of the parameter space is guided by the acquisition function $a(\mu(\mathbf{x}), s(\mathbf{x}))$ balancing the predicted functionality $\mu(\mathbf{x})$ and uncertainty $\sigma(\mathbf{x})$. This practically means that regions of high uncertainty may not be explored if the optimization deems there to be little chance that probing such areas will result in a high function value.

Here, we develop the GP-BO framework optimized for Scanning Probe Microscopy. All BO routines were implemented in *GPim*,[30] which is a home built software package designed for applications of GP and GP-based BO to image data and hyperspectral data. In GP-based BO, we start by choosing an appropriate GP prior, $\mathcal{GP}(f; m, K_f)$, over the true function, where *m* is a mean function (usually set to 0) and $K_f$ is a covariance function (kernel). The latter, as mentioned earlier, defines the strength and functional form of correlations between the values of the true function in the parameter space.[31] GPim provides a set of standard off-the-shelf kernels, including Gaussian radial basis function, Matern kernel, and Rational Quadratic kernel. Both exact GP and inducing points-based sparse[32] GP methods are available.

A single step of GP-based BO procedure then consists of i) obtaining the posterior distribution *via* Bayes rule given a set of function evaluations (observations); ii) deriving an acquisition function $\alpha(\mathbf{x})$ from the posterior; iii) finding the next query point(s) according to $\mathbf{x}_{n+1} = argmax(\alpha(\mathbf{x}_n))$; iv) performing measurements in the sampled point(s) and updating the posterior.

The choice of acquisition function is critical to the successful BO performance.[33] *GPim* provides three standard acquisition functions including confidence bound (CB), the probability of improvement (PI) and expected improvement (EI). However, the primary feature of the *GPim* package is the flexibility to design and implement custom acquisition function based on expected value and uncertainty and incorporate the other sources of information into the acquisition function (*e.g.* high-resolution topographic or piezoresponse data set). Furthermore, *GPim* allows the user to configure pathfinder functions, suggesting the optimal sequence of measurement locations from the almost-degenerate list of optima of the acquisition function. This approach in principle allows one to configure the measurement sequence given the peculiarity of the imaging system.

For the standard acquisition functions, the CB approach is simply a linear combination of GP mean, $\mu(\mathbf{x})$, and GP uncertainty, $\sigma(\mathbf{x})$, for each point,

$$\alpha_{CB} = a\mu(\mathbf{x}) + b\sigma(\mathbf{x}), \qquad (1)$$

Here the choice of coefficients controls a balance between exploration (minimization of uncertainty across the parameter space) and exploitation (maximization of the desired behavior) when selecting the next measurement point(s). Note that for $a = 1$ and $b > 0$ the CB becomes the well-known upper confidence bound acquisition function.[33] Another built-in option is a probability of improvement (POI) acquisition function[34], which tells how likely an improvement is and is defined as

$$\alpha_{PI} = \Phi\left(\frac{\mu(\mathbf{x})-y^+-\xi}{\sigma(\mathbf{x})}\right), \qquad (2)$$

where $\Phi$ is a standard normal cumulative distribution function, $y^+$ is the best observed value (for noisy data, we use a GP-predicted value) and $\xi$ determines a balance between exploration and exploitation during the optimization. Finally, the expected improvement (EI)[35] acquisition function can also account for the size of the improvement and is computed according to

$$\alpha_{EI} = (\mu(\mathbf{x}) - y^+)\Phi\left(\frac{\mu(\mathbf{x})-y^+-\xi}{\sigma(\mathbf{x})}\right) + \sigma(\mathbf{x})\phi\left(\frac{\mu(\mathbf{x})-y^+-\xi}{\sigma(\mathbf{x})}\right), \qquad (3)$$

when $\sigma > 0$ and vanishes otherwise. Here $\phi$ is the standard normal probability density function.

The derived acquisition function values are sorted in the descending order and by default the measurement coordinates corresponding to the first value (*i.e.* maximum) are chosen. In addition, to avoid getting locked in the same point, we introduce a short-term memory option, which controls a distance from the proposed measurement coordinate to the previous coordinates according to $\gamma^{t-1}d$ where $d$ and $\gamma$ are set by an operator and $t = 1$ for the coordinates at the previous step, $t = 2$ for the step before that, and so on.

Finally, although BO is a sequential algorithm (*i.e.* it outputs a single point at the end of each step), we found that in the context of the autonomous experiments it is often more efficient to have a batch of points as the output at each step. When this option is selected, by default the points in the batch are separated by a distance more or equal to GP kernel length scale at that step but can also be set manually (to be the same for all the steps).

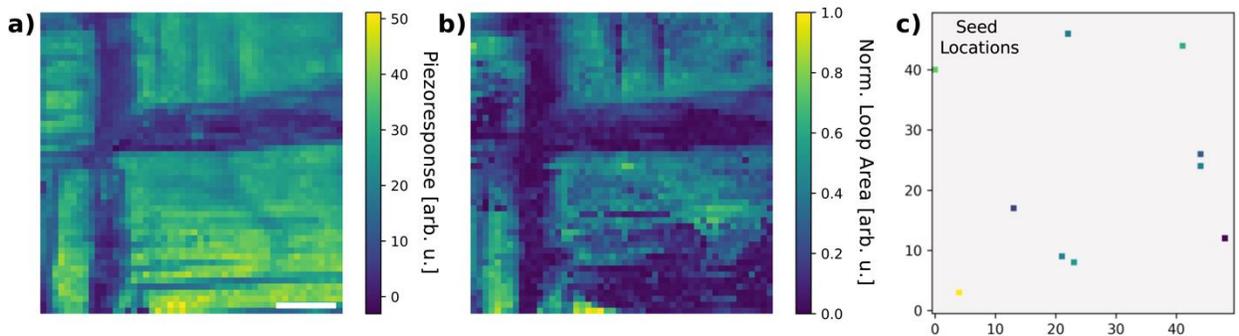

**Figure 1: Ground truth dataset for Gaussian Processing/Bayesian Optimization.** a) Piezoresponse spectroscopy grid (50x50) at 0V, b) normalized hysteresis loop area acquired at -15V to +15V, and c) initial random seed locations to start Gaussian Processing / Bayesian optimization.

We have systematically explored the BO AE on pre-acquired data collected *via* band excitation piezoresponse spectroscopy.[36] . Details of the acquisition are available elsewhere,[36, 37] but it should be noted that this technique requires fitting the cantilever response to a simple harmonic oscillator model to extract the amplitude, phase, frequency and Q-factor at every spectroscopic step. Hysteresis loops based on real and imaginary projections of the response were calculated, *i.e.* $f(V)_{real} = A\cos(\phi)$, $f(V)_{imag} = A\sin(\phi)$, and the loop area of the real and imaginary hysteresis loops $LA_{real}$ and $LA_{imag}$ were calculated. Then, the total loop area = $\sqrt{LA_{real}^2 + LA_{imag}^2}$ was determined. Note, this is important because the phase offset is not known *a priori*, which can result in challenges if only the real projection is utilized.

**Results and Discussion**

The piezoresponse map (50x50 grid) of PbTiO$_3$ at 0V is shown in Figure 1a, displaying a clear dense a-c domain structure with varying contrast ideal for exploring AE. The normalized hysteresis loop areas from -15V to +15V are shown in Figure 1b with distinct localized areas of maximum loop area or large electromechanical response (green-yellow regions), which are targeted for the BO AE (further details can be found in Supplementary Section). For initializing BO automated experiments, 10 seed locations are chosen at random providing a starting point for the exploration-exploitation, as shown in Figure 1c. Although we chose loop area in this instance, it should be noted that any feature of the spectra can be calculated and used, for instance, the coercive field, the remnant polarization, and so forth. Loop area is chosen specifically because it is not a tri*via*l property to estimate where it should be highest simply by inspection of the PFM image, and further, because it is important in the context of ferroelectric switching and energy loss.

The three standard acquisition functions implemented in BO automated experiments are shown in Figure 2, including EI (column 1-2), POI (column 3-4), and CB (column 5-6) acquisition functions. Note that here we are performing 'experiments' on the already collected data from Figure 1. Such tests are useful to understand the dependence of the process on hyperparameters that can then assist the user to set these parameters during real experiments. The exploration-exploitation tradeoff coefficient $\xi$ in Eq. (2) and Eq. (3) was set to 0.01 for both EI and POI. For CB, we used $a = 1$ and $b = 2$. For expected improvement, as the number of exploration steps are increased from 200-600 (rows 1-3), locations of enhanced electromechanical response (maximum loop area) are systematically explored resulting in clear similarities between the GP image reconstructions (columns 2,4, and 6) and ground truth image in Figure 1b. Similarly, the probability of improvement acquisition function targeted locations of enhanced response with the spatial variation in exploration steps seemingly localized. As for the confidence bound acquisition function, the initial steps explore slightly different regions while also staying rather localized. However, at 600 steps, all acquisition functions explore similar regions. It is important to note, the exploration steps displayed in Figure 2 are for a BO short-term memory parameters $d = 10$ and $\gamma = 0.8$, and a Radial Basis function (RBF) kernel length scale with minimal and maximal bounds set to [1, 5], respectively (the values are in image pixels). As such, these parameters influence the exploration-exploitation process.

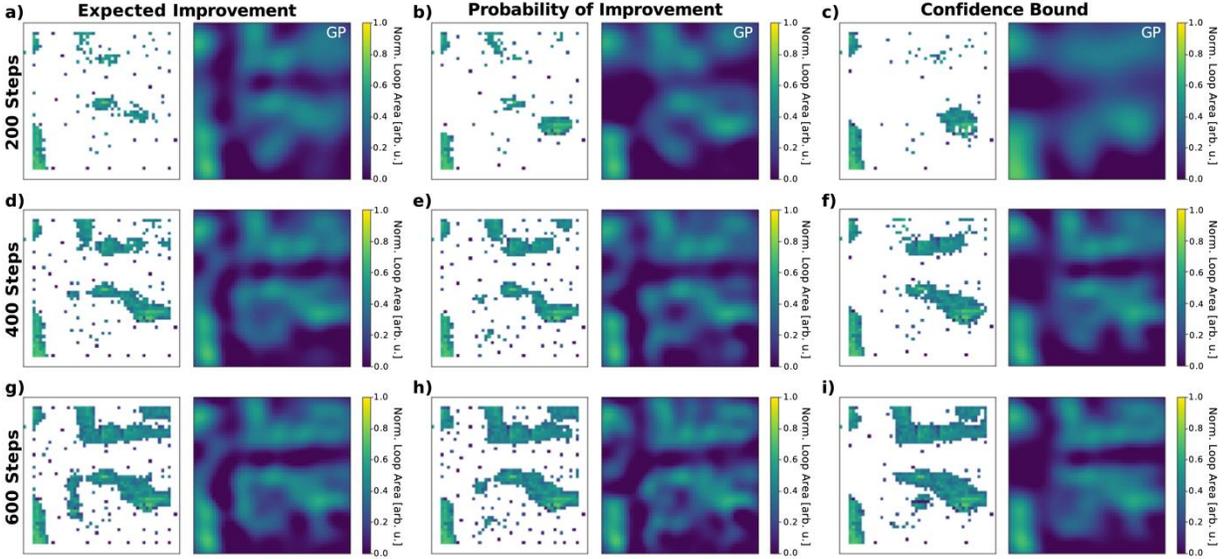

**Figure 2: Automated experiments with GP-BO for three different acquisition functions.** a-c) 200, d-f) 400, and g-i) 600 exploration steps. Columns 1-2, 3-4, and 5-6 are expected improvement, probability of improvement, and confidence bound acquisition functions, respectively. Sparse plots indicate locations explored based on the GP-BO predictions, while complete images show GP-based image reconstructions from sparse plots. Note, initial seed locations are shown in Figure 1c and are identical for all panels.

To explore the dependence of GP-BO algorithm on the initialization parameters, the short-term memory distance parameter and kernel length scale were varied allowing to go from the localized to global exploration regime (additional information in Supplementary Section), as shown in Figure 3. As expected, for all acquisition functions, small distance parameters and kernel length scales result in explorations tending to be somewhat localized until a certain number steps is acquired, where the subsequent steps result in a transition to a different location (Figure 3a,d,g). As the values of two parameters are increased, a broader exploration is observed, starting in the bottom left corner and branching out (Figure 3b,e,h) with majority of the larger electromechanical responses identified. Further, for completeness, we implement a large distance parameter with a small kernel length scale (Figure 3c,f,i), resulting in more sampling of the overall area, consequently locating more regions of larger enhanced response. As such, the choice of GP-BO initialization parameters and acquisition function distinctly effect the AE process, and should be chosen according to the spatial characteristics of the element of interest. Note that the choice of acquisition function does not appear to have a substantial effect for this case (see supplementary Figure S1).

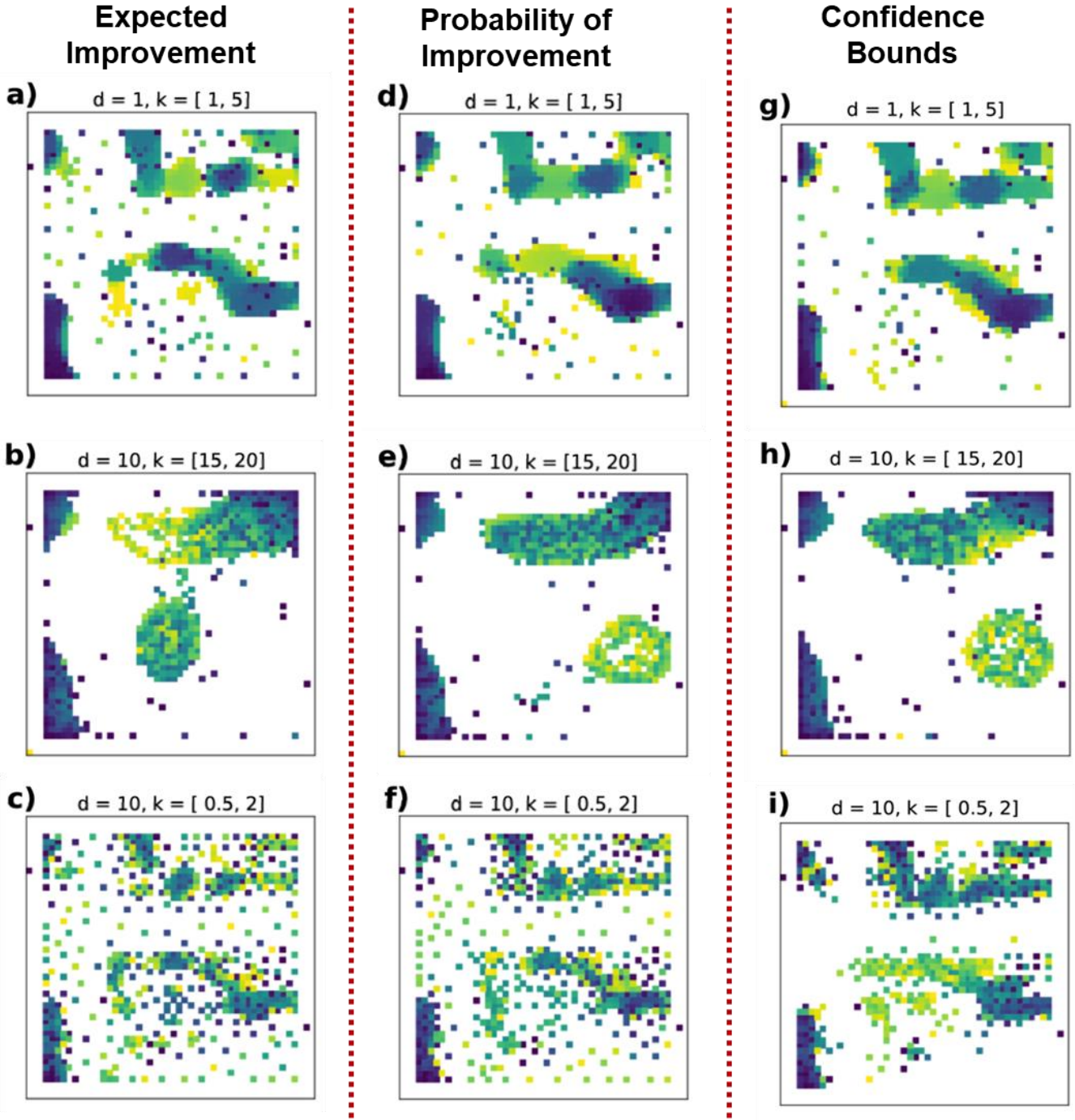

**Figure 3: GP-BO behavior for different optimization (hyper)parameters.** Exploration dependence on short-term memory distance parameter (d) and kernel length scale (k) for acquisition functions a-c) expected improvement, d-f) probability of improvement, and g-i) confidence bound. Step number is indicated by colorbar on the right

To implement the AE on the operational microscope, we combined in-house LabView-based scripts with National Instruments I/O hardware to control the tip positioning, bias waveform generation, and data acquisition, with a network link to a DGX-2 server with a bank of GPUs to enable the Bayesian optimization. Initially, as previously shown above, 10 randomly selected spatial locations were measured with band-excitation piezoresponse spectroscopy to obtain piezoresponse hysteresis loops. The calculated loop areas were used as the objective to maximize during the Bayesian optimization. Data was then communicated to a GPim script running on the NVIDIA's DGX-2 GPU server, for a list of the next 10 acquisition points. GP-BO was run for 500 iterations with the expected improvement acquisition function and an RBF kernel with length scale parameter's minimum/maximum bounds set to [2, 25] and the distance parameter d=13.5. To avoid edge effects, a border mask was used with a size of 2 px from all four sides, and the Bayesian optimizer excluded any points that are suggested for measurement within the mask, similar to the BO AE implemented on the pre-acquired data (above). File transfers to the DGX-2 were near instantaneous after each batch of hysteresis loops was acquired (<0.1s), and running the single GP-BO step on each batch took ~6 s on average. The acquisition time for each hysteresis loop was ~2.35 s. A schematic of the workflow used is shown in Fig. 4. The topography, vertical PFM amplitude and phase images are shown in Fig. 5(a-c), respectively, and the film contains a dense ferroelastic domain structure as explored previously.[38] Spectroscopy was performed at acquisition points across a 50x50 grid. A batch size of 10 was used, and the process was continued for a total of 40 steps, *i.e.* until 400 pixels worth of data was captured. Note that the kernel length size minima and maxima values were drawn from conservative inspection of Figure 5(b,c). The dense a/c domain regions would not be more than ~5 pixels across, but the larger a-domain might be perhaps 15 px across at its widest. To be more conservative, we therefore chose kernel length scale limits of [2,25].

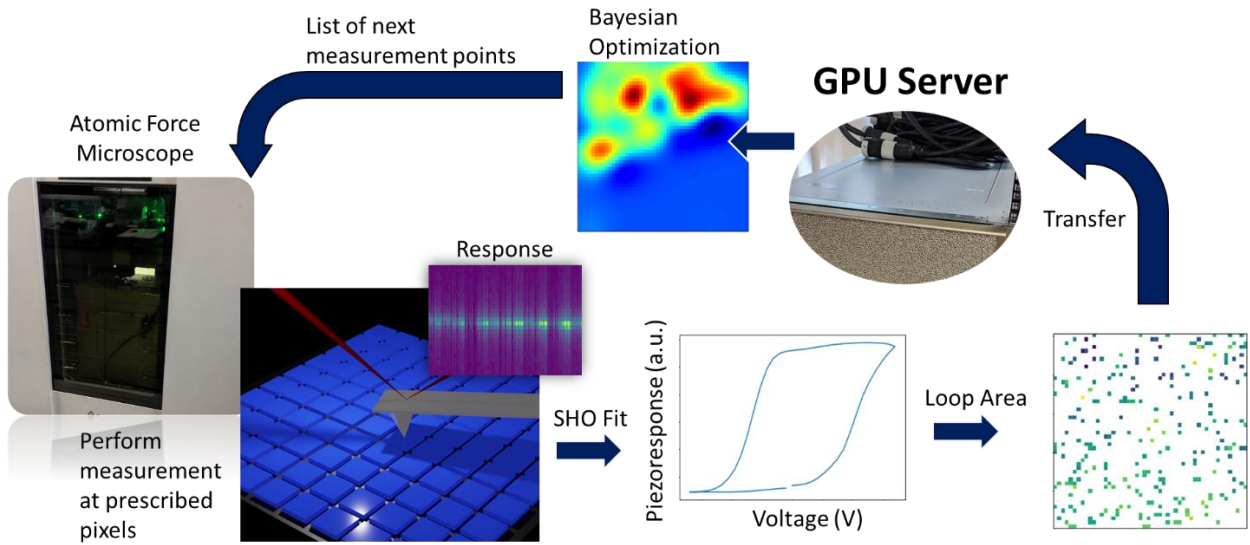

**Figure 4: Schematic of the experimental workflow.** The spectroscopy is performed on a list of points, the response is processed at the instrument to yield the loop area, after which it is then transferred to the DGX-2 system for GP-BO. This provides the list of measurement points to acquire next, which is ingested by the microscope software, and the process repeats.

The results of the automated experiment are shown in Figure 6, for 10, 20, 30, and 40 steps of the process. Between 30 and 40 GP steps, three distinct regions of high loop area are found, and the large a-domain cutting through the sample has also been found (and avoided). The full video of all the GP steps, to see the progression, is available in the supplementary video. It should be noted there is some drift of the spectroscopy results with respect to the initially captured PFM image. To confirm whether the areas predicted by the GP (after 400 pixels are measured) are indeed the areas with the highest response, we again performed a full spectroscopic acquisition in the standard experimental mode, *i.e.* capturing all pixels. This provides the ground truth, which is shown in Fig. 7(a). The final GP prediction after 40 iterations from the AE is shown for comparison in Fig. 7(b). The loop area has been normalized for convenience. The ground truth is again somewhat drifted from the AE experiment, but it is clear that there are primarily three clusters of points with high response (circled in Fig. 7(c)). Overlaying (and adjusting for drift) the GP prediction on this map, as in Fig. 7(d), reveals that the AE was able to detect two of these three clusters of points, but was unable to detect the third, on the middle-right of the image. Still, this is a good result that could be improved with a higher amount of exploration upfront or specific use of prior information to guide the initial stages of the exploration. In terms of efficiency, the AE experiment took 1180s, as compared to 5875s for the full acquisition without AE, *i.e.*, only 20% of the time.

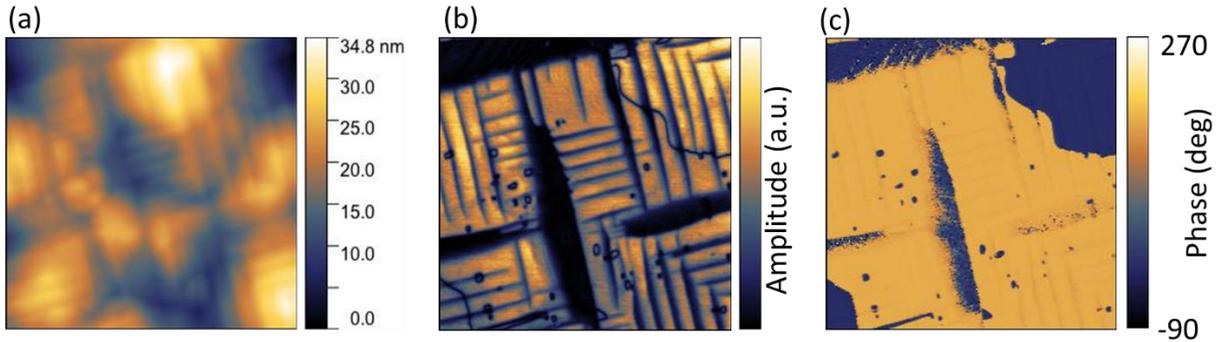

**Figure 5: Topography and PFM scan of PbTiO$_3$ thin film** (a) topography, Single frequency (b) Vertical PFM Amplitude, and (c) Vertical PFM Phase image of PbTiO$_3$ thin film (2um x 2um) showing a dense hierarchical ferroelastic domain structure.

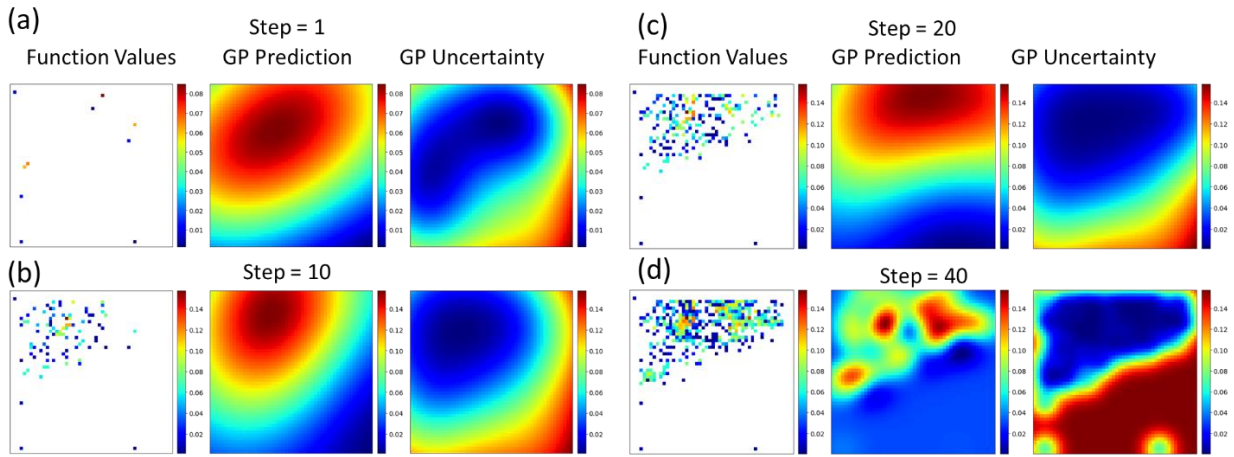

**Figure 6: Results of the automated experiment.** The function values (*i.e.*, the loop area measured), the GP prediction, and the GP uncertainty for (a) 1, (b) 10, (c) 20 and (d) 40 GP steps is shown. Over time the process exploits the region where high function values exist, but also learns to avoid the large a-domain cutting through the sample. The bottom region remains unexplored, because of the stochastic nature of the initialization. Area of scan is 2 um x 2 um.

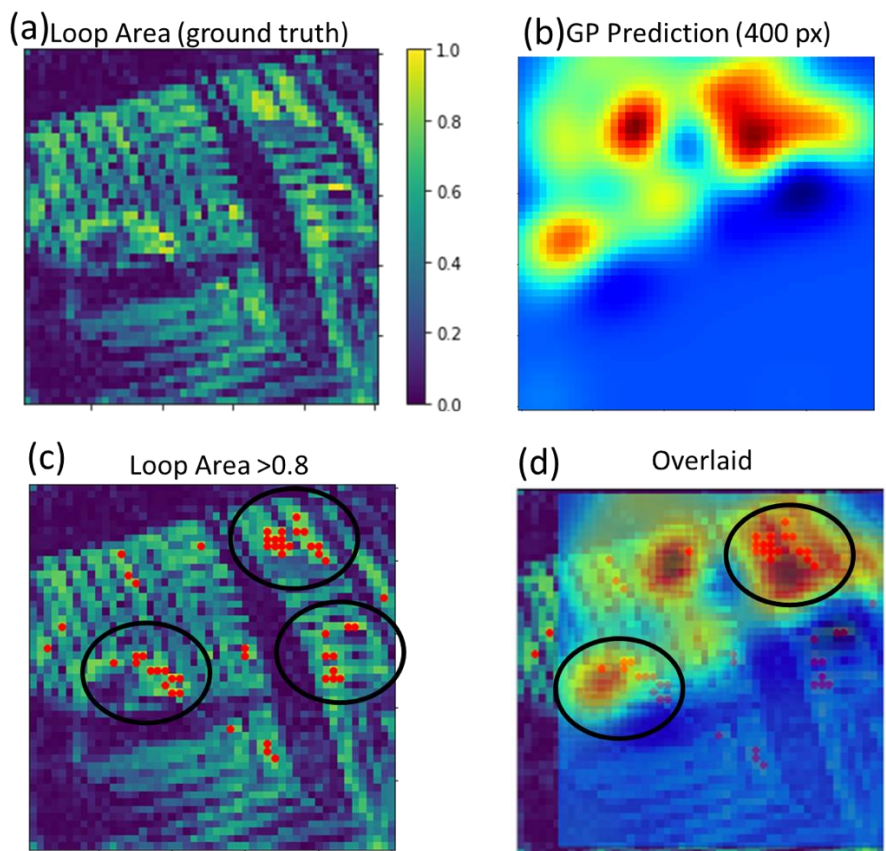

**Figure 7: Comparison of the AE results with 'ground truth'.** (a) The normalized loop area, *i.e.* ground truth. (b) GP Prediction after 40 GP steps (when 400 pixels have been measured). (c) Loop area with points higher than 0.8 marked in red. Three main clusters of higher values are found and are circled. When compared with the GP prediction, two of the three clusters are correctly predicted (d). Area of scan is 2 um x 2 um

**Prior Knowledge Incorporation**

One of the drawbacks of the approach outlined above are that it does not utilize prior knowledge of the domain structure, which is likely to be linked to the spectra themselves. For instance, a subset of the current authors have previously shown that autoencoders can be utilized to learn the mapping between imaging and spectral response.[39, 40] These studies, in addition to basic physics considerations, would suggest that incorporating the knowledge of the domain structure should be highly beneficial in optimizing the sampling process. To improve on the process, we utilized a secondary approach where in addition to utilizing the measured spectral values in the GP surrogate model, we also utilize the high-resolution imaging data itself as a secondary input channel, to aid predictions. This can be done by either simply inputting the raw pixel values of the image, or *via* inputting learned features from a convolutional neural network (CNN). The CNN is trained along with the GP model in the process and is a form of deep kernel learning. [41]We utilized the radial basis function (RBF) kernel, along with a simple convolutional neural network with 4 layers, with filter numbers of [1,2,4,8] respectively, and compare the technique against random search as well as one where only the image pixel values are utilized. Note that in this trial we only select a single pixel to sample next, as opposed to a batch.

In Figure 8 we plot the results comparing this method with the traditional GP surrogate modeling, applied to a dataset where the PFM image was acquired [Fig. 8(a)], and the spectroscopy was acquired subsequently with minimal drift [Fig. 8(b)]. The different methods used for the comparison are, 'random' (fully random search), 'position' (no prior knowledge from the image is used), 'image' (where the image pixel values are used directly as a secondary channel to the gaussian process regression (GP), and 'CNN', which uses the learned features as input to the GP, with the results are plotted in Fig. 8(c). It is evident that the 'CNN' method is superior for this particular trial run but note that there was run to run variability due to stochastic nature of initialization of the CNN weights. Indeed, the CNN requires training to learn the correlates between the PFM image and the loop area, which are likely to be somewhat complex[39] and dependent on both CNN parameter initializations and the initial pixels sampled. Once these features are learned, they can be used to improve the prediction, and it does appear that after ~500 pixels are sampled the CNN method is either comparable to, or superior, to the other methods over 10 runs. It should be noted that drift can render both the 'CNN' and 'image' method substantially worse than not using the prior information of the image at all. From a more practical standpoint, we note that implementation on the real microscope remains a challenge due to computational efficiency concerns, which will need to be improved and for which work is ongoing.

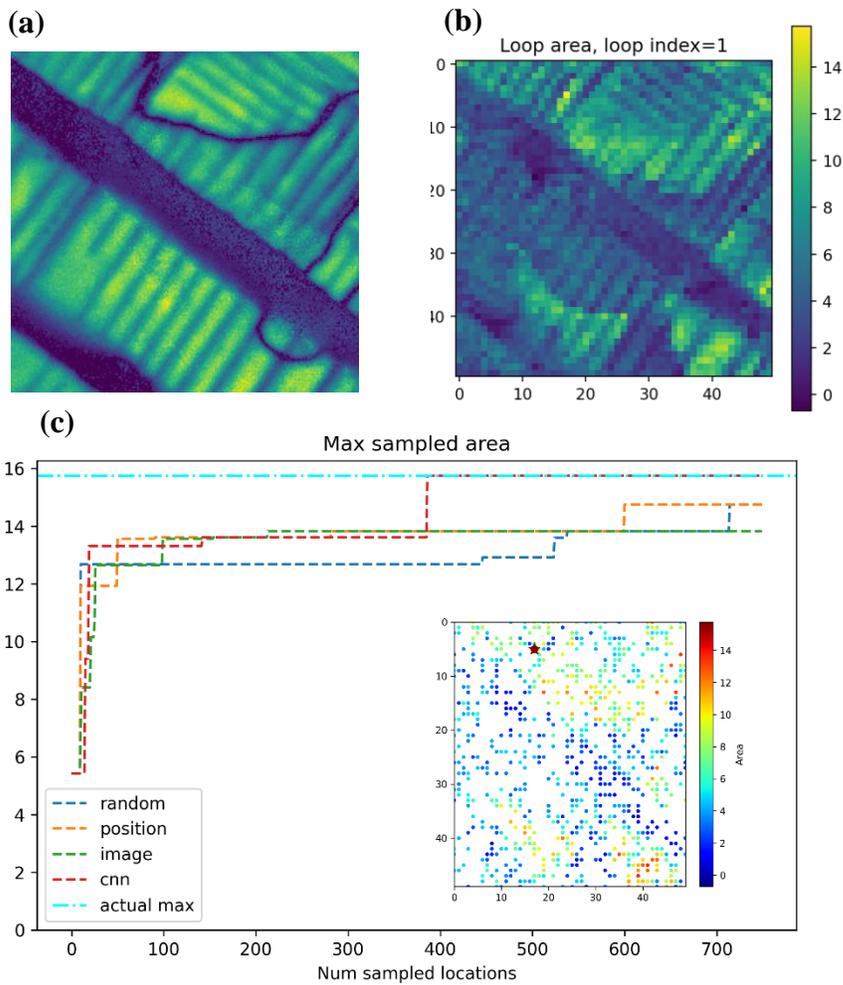

**Figure 8: Incorporating prior information in the form of a high-resolution PFM image. (a)** 2um x 2um Vertical Piezoresponse Force Microscopy amplitude image of the PbTiO$_3$ thin film. **(b)** Spectroscopy performed in the same region as (a), with the loop area calculated and plotted as a 2D map. Note the very minimal drift between (a) and (b). **(c)** Comparison of different surrogate models used to inform the Bayesian optimization process. (inset) Sampled pixels and their respective area, with the 'CNN' approach, and the pixel with highest area indicated by the star. The 'CNN' method was found to have similar or superior performance to the other methods after 500 sampled pixels in a sample of 10 trials.

**Conclusion**

We have developed a flexible Bayesian optimization workflow for the automated experiment in scanning probe microscopy and implemented it for the hysteresis loop measurements in Piezoresponse Force Microscopy. This approach allows problem-specific tuning of the BO workflow and operates in real time, the advance made possible by the incorporation of the edge computing based on a GPU server. Generally, the developed workflow is universal and can be adapted to other SPM and electron microscopy methods. We note that in the simplest form, GP-BO requires significant tuning that can be implemented *via* pre-acquired similar data. We further note that the efficiency of approach can be significantly improved by incorporation of the prior knowledge to initialize seed points. For example, in the case of the loop area maximization it can be expected that the initialization of the process should start at regions with high piezoelectric response from the initial piezoresponse scan (Figure 5). Alternatively, other schemes which take the gradient information from the piezoresponse map and weight them more heavily may also be considered, as it is expected that more information gain will occur in those regions than, for example, exploring within large, individual a- or c-domains. This also points to the need for specific domain expertise. Finally, once a CNN is trained once for one sample, it is likely to be highly beneficial for other experiments on the same sample, and likely potentially useful for other similar samples as a starting point. We conclude by noting that the utility of AE in SPM or any type of microscopy is conditional on the efficiency gains, and large gains will often require domain expertise and prior physics knowledge to be incorporated into the process.

**Acknowledgements:**
This research (GPim development, AE experimentation) was conducted at the Center for Nanophase Materials Sciences, which also provided support (RKV, MZ, SJ, SKV) and is a US DOE Office of Science User Facility. Simulated data analysis (KK) was supported by the U.S. Department of Energy (DOE), Office of Science, Basic Energy Sciences (BES), Materials Sciences and Engineering Division and was performed at the Oak Ridge National Laboratory's Center for Nanophase Materials Sciences (CNMS), a U.S. Department of Energy, Office of Science User Facility. This work (sample preparation) was partially supported by the JSPSKAKENHI Grant Nos. 15H04121, and 26220907 (H.F.).

**Methods/Experimental**

As a material system, we have chosen a 700 nm thick $PbTiO_3$ thin film grown by chemical vapor deposition on (001) $KTaO_3$ substrates with a $SrRuO_3$ conducting buffer layer, as reported by H. Morioka *et al.*[42, 43] The PFM was performed using an Oxford Instrument Asylum Research Cypher microscope with a National Instruments DAQ card and chassis, and operated with a LabView framework. All experiments were performed using Budget Sensor Multi75E-G Cr/Pt coated AFM probes (~3 N/m). All band excitation data was acquired with an AC excitation voltage of 2V applied to the tip.

*Financial Interests Statement*
The authors declare no financial interests.

**References**
1. Bonnell, D. A., Materials in Nanotechnology: New Structures, New Properties, New Complexity. *J. Vac. Sci. Technol. A* **2003,** *21* (5), S194-S206.
2. Gerber, C.; Lang, H. P., How the Doors to the Nanoworld Were Opened. *Nat. Nanotechnol.* **2006,** *1* (1), 3-5.
3. Pennycook, S. J.; Chisholm, M. F.; Lupini, A. R.; Varela, M.; van Benthem, K.; Borisevich, A. Y.; Oxley, M. P.; Luo, W.; Pantelides, S. T., Materials Applications of Aberration-Corrected Scanning Transmission Electron Microscopy. In *Advances in Imaging and Electron Physics, Vol 153*, Hawkes, P. W., Ed. Elsevier Academic Press Inc: San Diego, 2008; Vol. 153, pp 327-+.
4. Zhou, P. L.; Yu, H. B.; Shi, J. L.; Jiao, N. D.; Wang, Z. D.; Wang, Y. C.; Liu, L. Q., A Rapid and Automated Relocation Method of an Afm Probe for High-Resolution Imaging. *Nanotechnology* **2016,** *27* (39), 10.

5. Kalinin, S. V.; Borisevich, A.; Jesse, S., Fire up the Atom Forge. *Nature* **2016,** *539* (7630), 485-487.
6. Dyck, O.; Jesse, S.; Kalinin, S. V., A Self-Driving Microscope and the Atomic Forge. *MRS Bull.* **2019,** *44* (9), 669-670.
7. Krull, A.; Hirsch, P.; Rother, C.; Schiffrin, A.; Krull, C., Artificial-Intelligence-Driven Scanning Probe Microscopy. *Commun. Phys.* **2020,** *3* (1), 8.
8. Noack, M. M.; Yager, K. G.; Fukuto, M.; Doerk, G. S.; Li, R.; Sethian, J. A., A Kriging-Based Approach to Autonomous Experimentation with Applications to X-Ray Scattering. *Sci. Rep.* **2019,** *9* (1), 11809.
9. Noack, M. M.; Doerk, G. S.; Li, R.; Fukuto, M.; Yager, K. G., Advances in Kriging-Based Autonomous X-Ray Scattering Experiments. *Sci. Rep.* **2020,** *10* (1), 1325.
10. Scarborough, N. M.; Godaliyadda, G. M. D. P.; Ye, D. H.; Kissick, D. J.; Zhang, S.; Newman, J. A.; Sheedlo, M. J.; Chowdhury, A. U.; Fischetti, R. F.; Das, C.; Buzzard, G. T.; Bouman, C. A.; Simpson, G. J., Dynamic X-Ray Diffraction Sampling for Protein Crystal Positioning. *J. Synchrotron Radiat.* **2017,** *24* (1), 188-195.
11. Xing, H.; Zhao, B.; Wang, Y.; Zhang, X.; Ren, Y.; Yan, N.; Gao, T.; Li, J.; Zhang, L.; Wang, H., Rapid Construction of Fe–Co–Ni Composition-Phase Map by Combinatorial Materials Chip Approach. *ACS Combinatorial Science* **2018,** *20* (3), 127-131.
12. Kusne, A. G.; Gao, T.; Mehta, A.; Ke, L.; Nguyen, M. C.; Ho, K.-M.; Antropov, V.; Wang, C.-Z.; Kramer, M. J.; Long, C., On-the-Fly Machine-Learning for High-Throughput Experiments: Search for Rare-Earth-Free Permanent Magnets. *Sci. Rep.* **2014,** *4* (1), 1-7.
13. Henson, A. B.; Gromski, P. S.; Cronin, L., Designing Algorithms to Aid Discovery by Chemical Robots. *ACS central science* **2018,** *4* (7), 793-804.
14. Steiner, S.; Wolf, J.; Glatzel, S.; Andreou, A.; Granda, J. M.; Keenan, G.; Hinkley, T.; Aragon-Camarasa, G.; Kitson, P. J.; Angelone, D., Organic Synthesis in a Modular Robotic System Driven by a Chemical Programming Language. *Science* **2019,** *363* (6423).
15. Chan, E. P.; Lee, J.-H.; Chung, J. Y.; Stafford, C. M., An Automated Spin-Assisted Approach for Molecular Layer-by-Layer Assembly of Crosslinked Polymer Thin Films. *Review of Scientific Instruments* **2012,** *83* (11), 114102.
16. Noack, M. M.; Doerk, G. S.; Li, R.; Streit, J. K.; Vaia, R. A.; Yager, K. G.; Fukuto, M., Autonomous Materials Discovery Driven by Gaussian Process Regression with Inhomogeneous Measurement Noise and Anisotropic Kernels. *Sci. Rep.* **2020,** *10* (1), 1-16.
17. Nikolaev, P.; Hooper, D.; Webber, F.; Rao, R.; Decker, K.; Krein, M.; Poleski, J.; Barto, R.; Maruyama, B., Autonomy in Materials Research: A Case Study in Carbon Nanotube Growth. *npj Comp. Mater.* **2016,** *2* (1), 16031.
18. DeCost, B.; Hattrick-Simpers, J. R.; Trautt, Z.; Kusne, A. G.; Campo, E.; Green, M. L., Scientific Ai in Materials Science: A Path to a Sustainable and Scalable Paradigm. *Mach. learn.: sci. technol.* **2020**.
19. Boyce, B. L.; Uchic, M. D., Progress toward Autonomous Experimental Systems for Alloy Development. *MRS Bull.* **2019,** *44* (4), 273-280.
20. Ren, F.; Ward, L.; Williams, T.; Laws, K. J.; Wolverton, C.; Hattrick-Simpers, J.; Mehta, A., Accelerated Discovery of Metallic Glasses through Iteration of Machine Learning and High-Throughput Experiments. *Science advances* **2018,** *4* (4), eaaq1566.
21. Kelley, K. P.; Ren, Y.; Morozovska, A. N.; Eliseev, E. A.; Ehara, Y.; Funakubo, H.; Giamarchi, T.; Balke, N.; Vasudevan, R. K.; Cao, Y.; Jesse, S.; Kalinin, S. V., Dynamic Manipulation in Piezoresponse Force Microscopy: Creating Nonequilibrium Phases with Large Electromechanical Response. *ACS Nano* **2020,** *14* (8), 10569-10577.
22. Ovchinnikov, O. S.; Jesse, S.; Kalinin, S. V., Adaptive Probe Trajectory Scanning Probe Microscopy for Multiresolution Measurements of Interface Geometry. *Nanotechnology* **2009,** *20* (25).

23. Kelley, K. P.; Ziatdinov, M.; Collins, L.; Susner, M. A.; Vasudevan, R. K.; Balke, N.; Kalinin, S. V.; Jesse, S., Fast Scanning Probe Microscopy *Via* Machine Learning: Non-Rectangular Scans with Compressed Sensing and Gaussian Process Optimization. *Small* **2020,** *16* (37), 6.
24. Sang, X. H.; Lupini, A. R.; Unocic, R. R.; Chi, M. F.; Borisevich, A. Y.; Kalinin, S. V.; Endeve, E.; Archibald, R. K.; Jesse, S., Dynamic Scan Control in Stem: Spiral Scans. *Adv. Struct. Chem. Imag.* **2016,** *2*.
25. Jesse, S., He., Q, Lupinin, A.R., Leonard, D., Oxley, M.P., Ovchinnikov, O., Unocic, R.R., Tselev, A., Fuentes-Cabrera, M., Sumpter, B.G., Pennycook, S.J., Kalinin, S.V., and Borisevich, A.Y., Atomic-Level Sculpting of Crystalline Oxides: Toward Bulk Nanofabrication with Single Atomic Plane Precision. *Small* **2015**, 1.
26. Jesse, S.; Hudak, B. M.; Zarkadoula, E.; Song, J. M.; Maksov, A.; Fuentes-Cabrera, M.; Ganesh, P.; Kravchenko, I.; Snijders, P. C.; Lupini, A. R.; Borisevich, A. Y.; Kalinin, S. V., Direct Atomic Fabrication and Dopant Positioning in Si Using Electron Beams with Active Real-Time Image-Based Feedback. *Nanotechnology* **2018,** *29* (25).
27. Ziatdinov, M.; Kim, D.; Neumayer, S.; Vasudevan, R. K.; Collins, L.; Jesse, S.; Ahmadi, M.; Kalinin, S. V., Imaging Mechanism for Hyperspectral Scanning Probe Microscopy *Via* Gaussian Process Modelling. *npj Comp. Mater.* **2020,** *6* (1), 21.
28. Kalinin, S. V.; Ziatdinov, M.; Vasudevan, R. K., Guided Search for Desired Functional Responses *Via* Bayesian Optimization of Generative Model: Hysteresis Loop Shape Engineering in Ferroelectrics. *J. Appl. Phys.* **2020,** *128* (2), 8.
29. Kalinin, S. V.; Valleti, M.; Vasudevan, R. K.; Ziatdinov, M., Exploration of Lattice Hamiltonians for Functional and Structural Discovery *Via* Gaussian Process-Based Exploration–Exploitation. *J. Appl. Phys.* **2020,** *128* (16), 164304.
30. Ziatdinov, M., Gpim: Gaussian Processes and Bayesian Optimization for Images and Hyperspectral Data. *GitHub repository, https://git.io/JvK1s* **2020**.
31. Rasmussen, C. E.; Williams, C. K. I., *Gaussian Processes for Machine Learning (Adaptive Computation and Machine Learning)*. The MIT Press: Cambridge, Massachusetts, 2005.
32. Quiñonero-Candela, J.; Rasmussen, C. E., A Unifying View of Sparse Approximate Gaussian Process Regression. *J. Mach. Learn Res.* **2005,** *6* (Dec), 1939-1959.
33. Shahriari, B.; Swersky, K.; Wang, Z.; Adams, R. P.; Freitas, N. d., Taking the Human out of the Loop: A Review of Bayesian Optimization. *Proceedings of the IEEE* **2016,** *104* (1), 148-175.
34. Kushner, H. J., A New Method of Locating the Maximum Point of an Arbitrary Multipeak Curve in the Presence of Noise. **1964**.
35. Mockus, J.; Tiesis, V.; Zilinskas, A., The Application of Bayesian Methods for Seeking the Extremum. *Towards global optimization* **1978,** *2* (117-129), 2.
36. Jesse, S.; Maksymovych, P.; Kalinin, S. V., Rapid Multidimensional Data Acquisition in Scanning Probe Microscopy Applied to Local Polarization Dynamics and Voltage Dependent Contact Mechanics. *Appl. Phys. Lett.* **2008,** *93* (11), 112903.
37. Jesse, S.; Kalinin, S. V., Band Excitation in Scanning Probe Microscopy: Sines of Change. *J. Phys. D-Appl. Phys.* **2011,** *44* (46), 464006.
38. Morioka, H.; Yamada, T.; Tagantsev, A. K.; Ikariyama, R.; Nagasaki, T.; Kurosawa, T.; Funakubo, H., Suppressed Polar Distortion with Enhanced Curie Temperature in in-Plane 90 Degrees-Domain Structure of a-Axis Oriented Pbtio3 Film. *Appl. Phys. Lett.* **2015,** *106* (4), 5.
39. Kalinin, S. V.; Kelley, K.; Vasudevan, R. K.; Ziatdinov, M., Toward Decoding the Relationship between Domain Structure and Functionality in Ferroelectrics *Via* Hidden Latent Variables. *ACS Appl. Mater. Inter.* **2021,** *13* (1), 1693-1703.
40. Roccapriore, K. M.; Ziatdinov, M.; Cho, S. H.; Hachtel, J. A.; Kalinin, S. V., Predictability of Localized Plasmonic Responses in Nanoparticle Assemblies. *arXiv preprint arXiv:2009.09005* **2020**.

41.     Wilson, A. G.; Hu, Z.; Salakhutdinov, R.; Xing, E. P. In *Deep Kernel Learning*, Artificial intelligence and statistics, PMLR: 2016; pp 370-378.
42.     Li, L. L.; Cao, Y.; Somnath, S.; Yang, Y. D.; Jesse, S.; Ehara, Y.; Funakubo, H.; Chen, L. Q.; Kalinin, S. V.; Vasudevan, R. K., Direct Imaging of the Relaxation of Individual Ferroelectric Interfaces in a Tensile-Strained Film. *Adv Electron Mater* **2017,** *3* (4).
43.     Morioka, H.; Yamada, T.; Tagantsev, A. K.; Ikariyama, R.; Nagasaki, T.; Kurosawa, T.; Funakubo, H., Suppressed Polar Distortion with Enhanced Curie Temperature in in-Plane 90 Degrees-Domain Structure of a-Axis Oriented Pbtio3 Film. *Appl Phys Lett* **2015,** *106* (4).

*Supplementary Information*
**Autonomous Experiments in Scanning Probe Microscopy and Spectroscopy: Choosing Where to Explore Polarization Dynamics in Ferroelectrics**

Rama K. Vasudevan, Kyle Kelley, Jacob Hinkle, Hiroshi Funakubo,
Stephen Jesse, Sergei V. Kalinin, Maxim Ziatdinov

**Supplementary Figure 1**: Metrics for performance for Figure 2

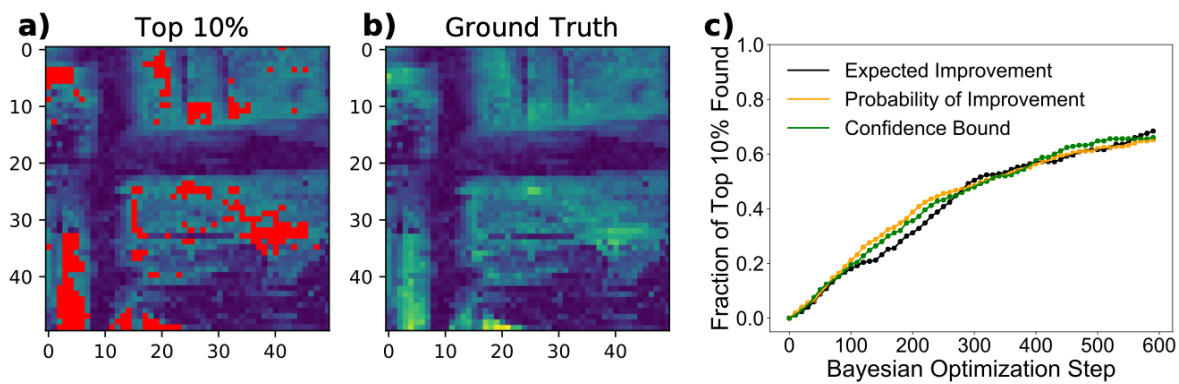

**Figure S1:** Time taken to find the top 10% of pixels as a metric of success for the example in Figure 2,3. **(a)** top 10% of pixels are highlighted in red. **(b)** Ground truth map**. (c)** Performance of the Bayesian optimization process for the three choices of acquisition functions.